\documentclass[a4paper,11pt]{article}
\usepackage{pos}
\usepackage{booktabs}
\usepackage[capitalise]{cleveref}

\def\eq#1{\begin{equation}#1\end{equation}}

\def\avg#1{{\langle{#1}\rangle}}

\def\dd{\mathrm{d}}
\def\eps{\varepsilon}

\title{Method for calculation of the beta exponent from the Heitler-Matthews model of hadronic air showers }
 \ShortTitle{Method for calculation of the beta exponent from the Heitler-Matthews model}

\author[a]{Kevin Almeida Cheminant }
\author*[a]{Dariusz G\'ora}
\author[a]{Nataliia Borodai}
\author[b]{Ralph Engel}
\author[b]{Tanguy Pierog}
\author[a]{Jan  P\c{e}kala}
\author[b]{Markus Roth}
\author[b]{Jarosław   Stasielak}
\author[b]{Michael Unger}
\author[b]{Darko  Veberi\v{c}}
\author[a]{Henryk   Wilczy\'nski}

\affiliation[a]{Institute of Nuclear Physics PAN, Radzikowskiego 152, Krakow, Poland}

\affiliation[b]{ Karlsruhe Institute of Technology (KIT),
Institute for Astroparticle Physics, Karlsruhe, Germany }

\emailAdd{Dariusz.Gora@ifj.edu.pl}

\abstract{The number of muons in an air shower is a strong indicator of the mass of the primary particle and increases with a small power of the cosmic ray mass by the $\beta$-exponent, $N_{\mu} \sim A^{(1-\beta)}$. This behaviour can be explained in terms of the Heitler-Matthews model of hadronic air showers. In this paper, we present a method for calculating $\beta$ from the Heitler-Matthews model. The method has been successfully verified with a series of simulated events observed by the Pierre Auger Observatory at $10^{19}$ eV. To follow real measurements of the mass composition at this energy, the generated sample consists of a certain fraction of events produced with p, He, N and Fe primary energies. Since hadronic interactions at the highest energies can differ from those observed at energies reached by terrestrial accelerators, we generate a mock data set with $\beta =0.92$ (the canonical value) and $\beta =0.96$ (a more exotic scenario). The method can be applied to measured events to determine the muon signal for each primary particle as well as the muon scaling factor and the $\beta$-exponent. Determining the $\beta$-exponent can effectively constrain the parameters that govern hadronic interactions and help solve the so-called muon problem, where hadronic interaction models predict too few muons relative to observed events. In this paper, we lay the foundation for the future analysis of measured data from the Pierre Auger Observatory with a simulation study.}

\FullConference{%
The 38th International Cosmic Ray Conference (ICRC2023)\\
  26 July -- 3 August, 2023\\
  Nagoya, Japan}

\begin{document}
\maketitle
\vspace{-0.4cm}
\section{Introduction}
\vspace{-0.3cm}
Simulations of extensive air showers using current hadronic interaction models predict too small numbers of muons compared to events observed in the air-shower experiments, which is known as the muon-deficit problem.
To study the muon-deficit we use the top-down (TD) method~\cite{augerTD,poro,goraicrc,physicascripta} -- This chain of simulations and reconstructions enable us to calculate signals in the fluorescence (FD) and surface detectors (SD) of cosmic ray experiments like for example the Pierre Auger Observatory or Telescope Array.
For each observed hybrid shower~\footnote{Hybrid event is seen simultaneously by the SD and FD detector.}, starting with a large number of simulated air showers with varying initial conditions, we select the one which has a longitudinal profile most similar to the profile of the observed shower (the reference profile).
As a result of the simulation-reconstruction chain we get an event, with complete information about the distributions of the signals in the detectors (including information on the specific components that contribute to these signals) -- these signals can then be compared with their reference counterparts.
Since the results of the simulations depend on the properties of the hadronic interaction models that are included in the simulation software, by comparing the simulations with corresponding observational results we should be able to verify these models at energies exceeding those available in any man-made accelerators.
We expect that we will gain new information, which will enable improvement of the interaction models, and in this way we are also able to reduce the discrepancy between the observations and simulations~\cite{augerTD,physicascripta}.

In this note the  method is proposed to  calculate the $\beta$-exponent from
the Heitler-Matthews model~\cite{mat} by including also the muon deficit-problem. The idea of the method is to find the set of muon rescaling parameters $\epsilon_k$ for different primaries $k$, which is a function of only two parameters: $\epsilon_\text{p}$ and $\Delta\beta$. These two parameters indicate how much we need to scale the proton signal ($\epsilon_\text{p}$ term) and by
how much to modify the $\beta$-exponent ($\Delta\beta$) in the Heitler-Matthews formula in order to match the observed numbers of muons in data and in simulations.
The method requires that  the first two  moments of the individual so-called $z_k$-distributions (our model) and overall $z$-distribution (the measured observable) are matched. In addition we require that the $\epsilon_k$ parameters should follow the Heitler-Mathews progression.
The $z_k$-distribution is essentially the difference between the total signal at 1000\,m of a real hybrid event and of the total signal at 1000\,m of the Monte Carlo (MC) dataset.
In other words, the method  tells us by how much each individual $z_k$-distribution must be shifted, rescaled and then, weighted and summed, in order to retrieve the $z$-distribution.
In TD-analysis we have the \emph{input dataset}, which are real or mock hybrid  events, and the \emph{matched dataset}, which is produced via \textsc{Conex}/\textsc{Corsika} Monte-Carlo simulations~\cite{heck}.
The input dataset contains $N$ events and the events will be indexed as $n=1,\ldots,N$.
The multiple profile-matched MC events simulated with primary $k$ corresponding to an input event $n$ are indexed with $i=1,\ldots,M$ and are thus denoted with the triplet subscript $nki$~\footnote{All $S^\text{x}$ symbols will be referring to the signal at 1000\,m from the shower core so that the 1000 subscript can be dropped entirely.
The signals at 1000\,m for the input dataset will have no decorations, i.e.\ just $S$, and the signals from the matched dataset will be denoted with $\bar{S}$.}.

\section{Two-parameter nonlinear scaling model}
\vspace{-0.2cm}
Observations of air showers, and also simulations, demonstrated that the number of muons $N_\upmu$ grows almost linearly with the shower energy $E$, and it also increases with a small power of the primary mass $A_k$.
These relations can be reproduced in the framework of the Heitler-Matthews model of hadronic air showers~\cite{mat}.
This model predicts
\eq{
N_\upmu^{A_k} =
  A_k\,\left(\frac{E/A_k}{\epsilon_\text{c}^\pi}\right)^\beta,
\label{HMmodel}
}
where $\beta\approx0.9$.
More precisely, MC simulations yield $\beta^\text{mc}=0.927\pm0.002$ for \textsc{Epos-LHC} and $\beta^\text{mc}=0.925\pm0.002$ for \textsc{QGSJetII-04}~\cite{15}.
For any fixed energy \cref{HMmodel} describes how the muon number depends on the primary mass: $N_\upmu^{A_k}=N_\upmu^\text{p}\,A^{1-\beta}_k$.\footnote{the $N_\upmu^\text{p}$ is the number of muons  in proton shower; $\epsilon_\text{c}^\pi$ is the critical energy at which pions decay into muons}
Simulations have shown that muon number depends on various properties of hadronic interactions (e.g.\ multiplicity, charge ratio, baryon anti-baryon pair production)~\cite{14}.
Therefore, estimating the $\beta$-exponent from data would be helpful in constraining the parameters of hadronic interactions and improving the accuracy of models.
On the other hand results obtained from the Pierre Auger Observatory and other leading cosmic ray experiments indicate that simulations using LHC-tuned hadronic interaction models underestimate the number of muons in extensive air showers compared to experimental data.
To account for this effect, we can formulate a scaling ansatz in \cref{HMmodel} by:
\eq{
N_\upmu^{A_k} = \bar{N}_\upmu^{A_k} \, A^{1-\beta}_k \, e^{\eps_\text{p}} \, A_k^{-\Delta\beta}.
\label{hmeq}
}
where the scaling factor can be defined as: $r_{\upmu,k} :=
  1 + \eps_k := e^{\eps_\text{p}} \, A_k^{-\Delta\beta} = \exp(\eps_\text{p} - \Delta\beta\ln A_k).$
%As we mentioned  above these two parameters indicate how much we need to scale the proton signal ($\eps_\text{p}$ term) and by how much to modify the $\beta$ exponent ($\Delta\beta$) in the Heitler-Matthews formula~\cite{mat} in order to match the observed numbers of muons in data and in simulations.
\begin{figure}
\centering
\includegraphics[width=0.55\textwidth]{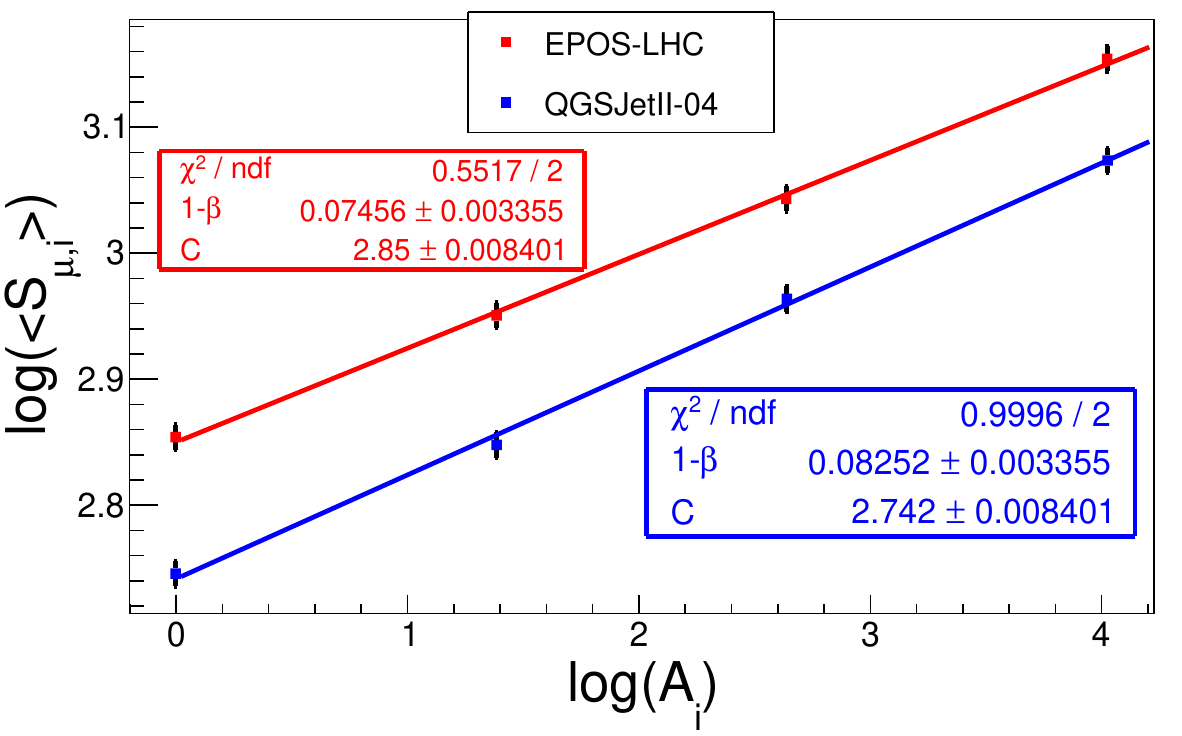}
\caption{Average logarithm of the muon signal for \textsc{Epos-LHC} and \textsc{QGSJetII-04}.
Solid lines are fits of the function $\bar{S}^\upmu_k=\text{const}\,A_k^{1-\beta}$ to the TD simulation.
From the fit, we obtain $\beta=0.925\pm0.003$ for \textsc{Epos-LHC} (red line), and $\beta=0.918\pm0.003$ for \textsc{QGSJetII-04} (blue line).}
\label{fig:beta-muons-mc}
\vspace{-0.3cm}
\end{figure}
Thus, having MC values of the $\beta^\text{mc}$ for the hadron interaction model and the value of the parameter $\Delta\beta$, we can calculate the $\beta$ exponent from $\beta =\beta^\text{mc} + \Delta\beta$.

In the context of this work, this then corresponds to saying that the number of muons $N_\upmu^{A_k}$ in the input dataset is proportional to the muon number $\bar{N}_\upmu^{A_k}$ in the matched dataset, with the usual Matthews-Heitler progression with mass $A_k$, but with a slight scaling $1+\eps_\text{p}$ and modification $\Delta\beta$.
In this work, the input dataset is constructed from \textsc{Epos-LHC} simulations (mock dataset) and is built by taking MC simulations from the TD simulation chain obtained with \textsc{Epos-LHC} around $10^{18}$\,eV.
The matched dataset is a dataset from \textsc{QGSJetII-04} simulations.
Details regarding these two datasets can be also found in Ref.~\cite{physicascripta}.

Since these simulations were performed for p, He, N, and Fe primaries for both \textsc{Epos-LHC} and \textsc{QGSJetII-04}, we can plot the evolution of the average muon signal as a function of the primary mass for both hadronic models, as shown in \cref{fig:beta-muons-mc}.
Since \textsc{QGSJetII-04} has, on average, fewer muons than \textsc{Epos-LHC}, one can imagine that the muon problem can be recreated by comparing the two hadronic models. Therefore, we can try in this work and figure what is the best way to rescale \textsc{QGSJetII-04} in order to match the muon signal of the mock dataset built with \textsc{Epos-LHC}.

From \cref{fig:beta-muons-mc} we can also see, that the average muon signal increases as a function of the primary mass.
As expected, both considered hadronic models display a similar ratio with the average about $r^\text{mc}_\text{true}=\bar{S}^\upmu_\text{epos}/\bar{S}^\upmu_\text{qgsjet}=1.10\pm0.04$; upon closer examination we also see that larger rescaling is needed  for protons ($1.12\pm0.03$) than for iron ($1.08\pm0.03$).
In \cref{fig:beta-muons-mc} we show also the linear fit  to the MC muon signal from \textsc{Epos-LHC} and \textsc{QGSJetII-04}, motivated by the Heitler-Matthews model.
The calculated value of $\beta^\text{mc}$ from the fit is about 0.92, so it is  pretty close to the values from Ref.~\cite{15}.
This cross-check of $\beta$-calculation is a validation of our TD simulations.

\begin{figure}
\centering
\def\w{0.5}
\includegraphics[width=\w\textwidth]{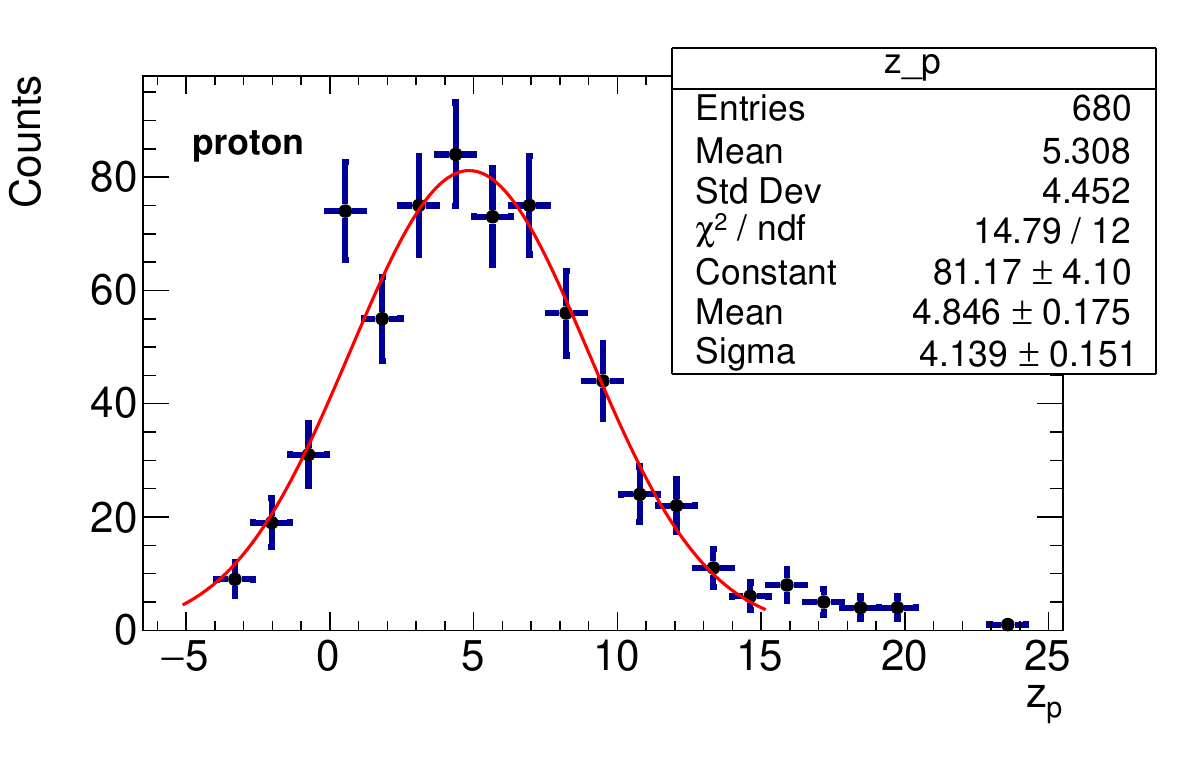}\hfill
%\includegraphics[width=\w\textwidth]{z_helium2}
%\\
%\includegraphics[width=\w\textwidth]{z_nitrogen2}\hfill
\includegraphics[width=\w\textwidth]{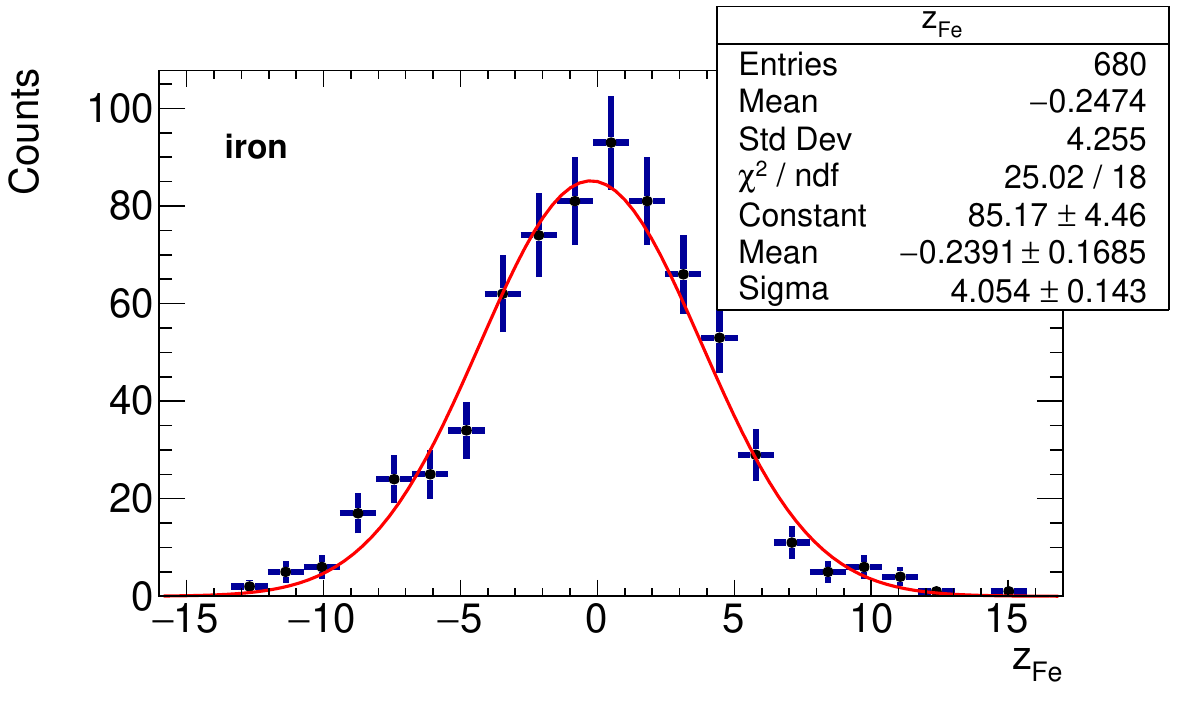}
\caption{The $z_k$-distributions for stations at 1000\,m from the shower core, from TD simulations at energy $10^{19}$\,eV for  proton (left)  and iron (right) induced air showers simulated with \textsc{Epos-LHC} and \textsc{QGSJetII-04} for the so-called mock dataset, see Ref.~\cite{physicascripta} for more details.}
\label{fig:individual-z-distribution}
\end{figure}
\section{Fitting the $z$-histogram }
\vspace{-0.2cm}
The mean signal $\avg{S}$ of the input dataset is the sum of the mean electromagnetic (em) and muonic components
\eq{
\avg{S} =
  \sum_k f_k \, \avg{S}_k =
  \sum_k f_k \left(\avg{S^\text{em}}_k + \avg{S^\upmu}_k\right) =
  \avg{S^\text{em}} + \avg{S^\upmu},
}
where $\avg{\cdot}_k$ denotes a mean within a given primary class $k$.
Note that for the input dataset the averages for given $k$ are not really observable, but it is clear that a sum over the composition fractions $f_k$ gives then the average in the whole input dataset, a quantity which is fully available.
Equivalently, for the mean signal $\avg{\bar{S}}$ in the matched dataset, where the quantities are known for various primary groups $k$, we can explicitly write
\eq{
\avg{\bar{S}} =
  \sum_k f_k \, \avg{\bar{S}}_k =
  \sum_k f_k \left(\avg{\bar{S}^\text{em}}_k + \avg{\bar{S}^\upmu}_k\right) =
  \avg{\bar{S}^\text{em}} + \avg{\bar{S}^\upmu},
}
where $\avg{\bar{S}}_k:=(\sum_n^N\sum_i^{M_{nk}}\bar{S}_{nki})/\sum_n^N M_{nk}$ is the signal $\bar{S}_{nki}$ of the matched dataset averaged over all $n$ and $i$ for a given $k$.

Since we assume a perfect matching of the longitudinal profile and thus the EM component of the signal, all the $\bar{S}^\text{em}_{nki}$ are very close or identical to the corresponding input events with signals $S^\text{em}_n$.
The mean difference $\Delta S$ of the signals in the two datasets thus only depends on the muonic part
\eq{
\Delta S :=
  \avg{S} - \avg{\bar{S}} =
  \sum_k f_k \left(\avg{S}_k - \avg{\bar{S}}_k\right) =
  \sum_k f_k \left(\avg{S^\upmu}_k - \avg{\bar{S}^\upmu}_k\right) =
  \avg{S^\upmu} - \avg{\bar{S}^\upmu} =
  \Delta S^\upmu.
\label{delta_s_s_mu}
}
The mean muonic signals $\avg{S^\upmu}_k$ of the primary $k$ in the input data can be obtained by rescaling the muonic signals $\avg{\bar{S}^\upmu}_k$ in the matched dataset with corresponding scaling factors $1+\eps_k$,
\eq{
\avg{S^\upmu}_k = (1 + \eps_k) \, \avg{\bar{S}^\upmu}_k.
\label{scaling}
}
With this scaling we can simplify the difference $\Delta S$ from \cref{delta_s_s_mu} into
\eq{
\Delta S^\upmu = \sum_k f_k \, \eps_k \, \avg{\bar{S}^\upmu}_k.
\label{delta_s_mu}
}
On the other hand, as it is clear from \cref{delta_s_s_mu}, $\Delta S\equiv\Delta S_\upmu$.
The third term of \cref{delta_s_s_mu} can be rewritten as
\eq{
\sum_k f_k \left(\avg{S}_k - \avg{\bar{S}}_k\right) =
  \avg{S} - \sum_k f_k \, \avg{\bar{S}}_k,
\label{delta_s_summed}
}
so that we can define for each event $n$ and match $i$ an observable
\eq{
z_{ni} = S_n - \sum_k f_k \, \bar{S}_{nki}.
\label{z}
}
Equivalently, based on \cref{delta_s_mu} we can define a scaling-dependent quantity
\eq{
\bar{z}_{ni} =
  \sum_k f_k \, \eps_k \, \bar{S}^\upmu_{nki} =
  \sum_k f_k \, \eps_k \, g_k(\theta_n) \, \bar{S}_{nki},
\label{z_bar}
}
where $\bar{S}^\upmu_{nki}$ is obtained either directly from the MC events or, like here, by using a factor $g$ from Universality, $\bar{S}^\upmu_{nki}=g_k(\theta_n)\,\bar{S}_{nki}$.The average muon signal as a fraction of the total signal at the ground, $g_{k}(\theta_n)$ has been calculated in our previous analyses, see for example~\cite{physicascripta}~\footnote{It is worth mentioning that this fraction depends on the shower zenith angle and the type of the primary cosmic ray, and only slightly on different hadronic interaction models~\cite{10}.}.

\begin{figure}
\centering
\def\h{0.36}
\includegraphics[height=\h\textwidth]{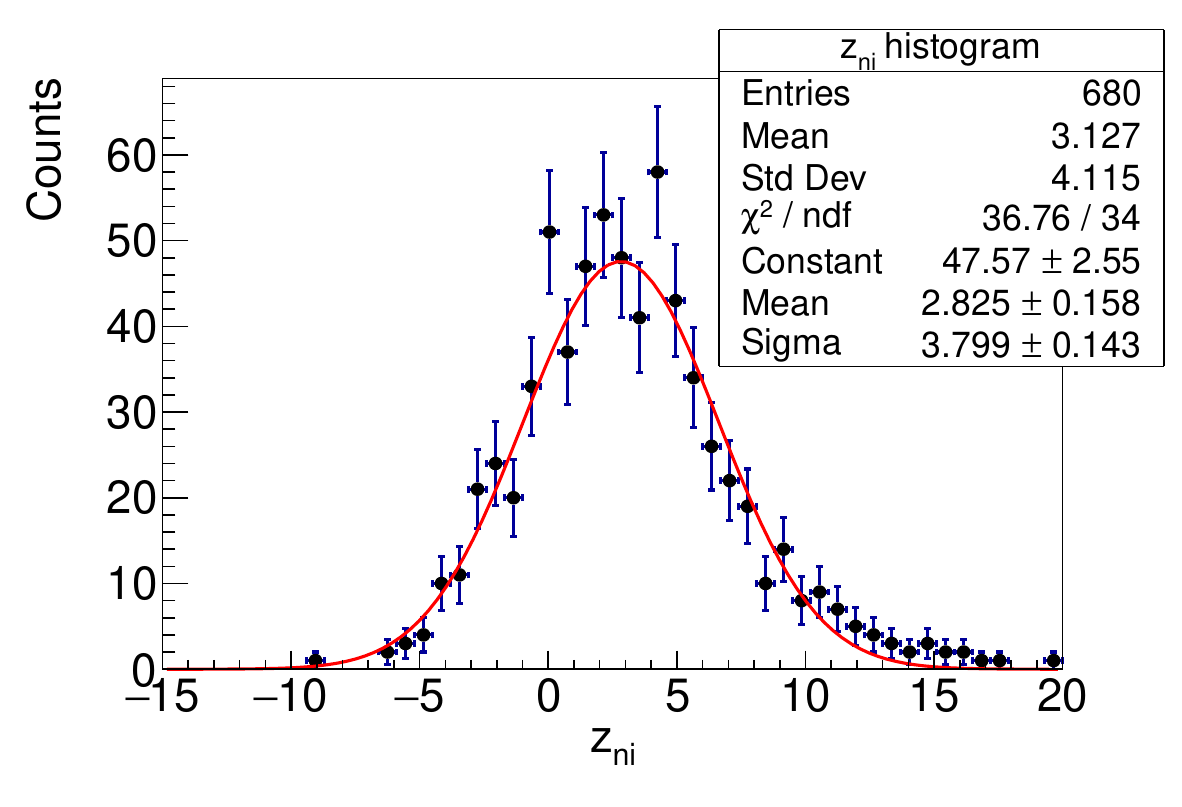}\hfill
\includegraphics[height=\h\textwidth]{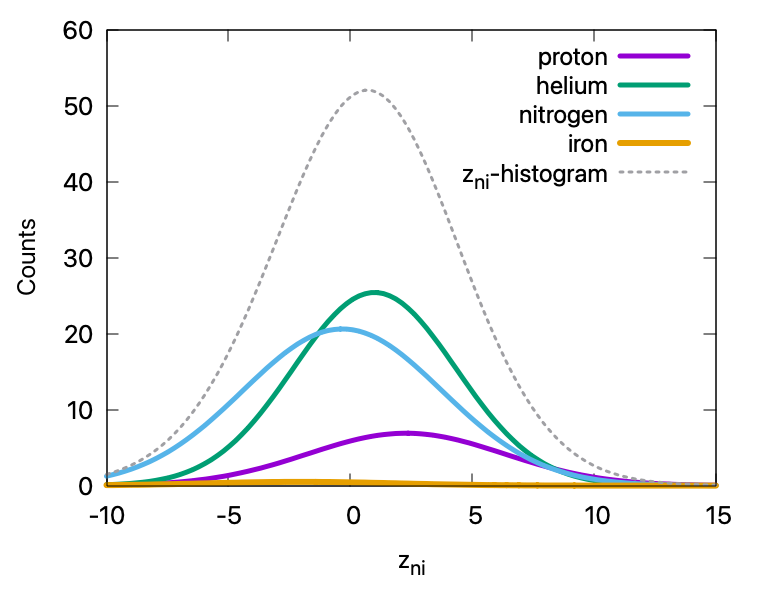}
\caption{(Left): The  $z_{ni}$-distribution as described by \cref{z} with $f_\text{p}=0.15$, $f_\text{He}=0.38$, $f_\text{N}=0.46$, and $f_\text{Fe}=0.01$  for mock dataset.
Since we have 68 mock events (\textsc{Epos-LHC}) and 10 \textsc{QGSJetII-04} events associated to each of the mock events we have 680 events contained in this histogram.
The distribution is fitted (red line) with a Gaussian function in order to get its mean $\avg{z_{ni}}=2.825\pm0.16$ and the standard deviation $\sigma(z_{ni})=3.80\pm0.14$.
(Right): Sketch showing the idea of the method i.e.\ each $z_k$-distribution must be shifted, rescaled, and then weighted and summed, in order to retrieve the $z_{ni}$-histogram.}
\label{fig:z-distribution}
\end{figure}

For each event $n$ and $i$ we can also define a variable $z_{nki}=S_{nki}-\bar{S}_{nki}$, which is a simple difference between the total signal for data and MC, for given primary $k$.
In \cref{fig:individual-z-distribution} we show corresponding distributions of this variable for the considered primaries obtained from TD simulations with \textsc{Epos-LHC} and \textsc{QGSJetII-04} (for simplicity we use notation $z_k$ for each individual histogram). As we can see from \cref{fig:individual-z-distribution} for the considered number of events, the corresponding $z_k$-distribution can be quite well described by a Gaussian function, the fit to histograms gives $\chi^2/\text{ndf}\approx1.5$.
From the fit for individual distributions we can get the mean value of signal difference $\avg{z_{k}}$ and the corresponding standard deviation $\sigma(z_k)$.
These variables can be used to define the probability density function (PDFs) for each primary $k$, which is given by
\eq{
P_{k}(z_k,\sigma(z_k)) =
  \frac{1}{\sqrt{2\pi}\sigma(z_k)}
  \exp\left[-\frac{(z_{nki} - \avg{z_k})^2}{2\sigma^2(z_k)}\right],
}
where again index $k$ spans over different primaries. 
 
Note, that according to \cref{z_bar} the mean position of $z_k$-distribution should be connected with an average ground muon signal expected for given primary.
However, such conversion is possible, if we already know proportionality constants i.e.\ scaling factors $\eps_k$.
In other words, if we plot rescaled distribution shown in \cref{fig:individual-z-distribution} in $\avg{S^\upmu}$ phase-space, the means of such distributions should give us average muon signals on the ground for considered masses.
Moreover, we should expect from physics of extensive air showers that position of mean for lighter element should be smaller that for heavier element i.e.\ $\avg{S^\upmu}_\text{p} < \avg{S^\upmu}_\text{He} < \avg{S^\upmu}_\text{N} < \avg{S^\upmu}_\text{Fe}$.
Based on the Heitler-Matthews model it is also expected, that logarithm of the muon signal should increase linearly with logarithm of the primary mass, therefore corresponding linearity conditions were introduced by using two-parameter scaling model $\eps_k$. 

In order to find $\eps_k$ and thus to convert the mean of $z_k$-distribution to $S^\upmu$ phase-space, we can use the Minuit minimization, where the fitted function is a combination of four Gaussian PDFs, which have the form
\eq{
F(\vec{\eps}, A_\text{mpl}) =
  A_\text{mpl}
  \sum_k f_k \, \frac{1}{\sqrt{2\pi}\sigma(z_k)}
    \exp\left[-\frac{(z_{nik} - {\color{red} \eps_k \avg{\bar{S}^\upmu}_k)^2}}{2\sigma^2(z_k)}\right],
\label{fit-final} 
}
where $\eps_k=e^{\eps_p-\Delta\beta\ln A_k}-1$ and $\text{const}=A_\text{mpl}$. The $f_k$ is fraction of $N=68$ pure mass samples and $const$ gives possibility to rescale the normalized individual $z_k$-distribution to overall $z_{ni}$-histogram.
 
In this way from the Gaussian fit given by \cref{fit-final} to overall $z_{ni}$-histogram, correction factors $\eps_k$ and $\Delta\beta$ for hadronic models can be calculated.
In other words, \cref{fit-final} tells us by how much each $z_k$-distribution must be shifted, rescaled, and then weighted and summed, in order to retrieve the $z_{ni}$-distribution and also its first and second moments, see also \cref{fig:z-distribution}.

\begin{figure}
\centering
\def\h{0.28}
\includegraphics[width=0.49\textwidth, height=\h\textwidth]{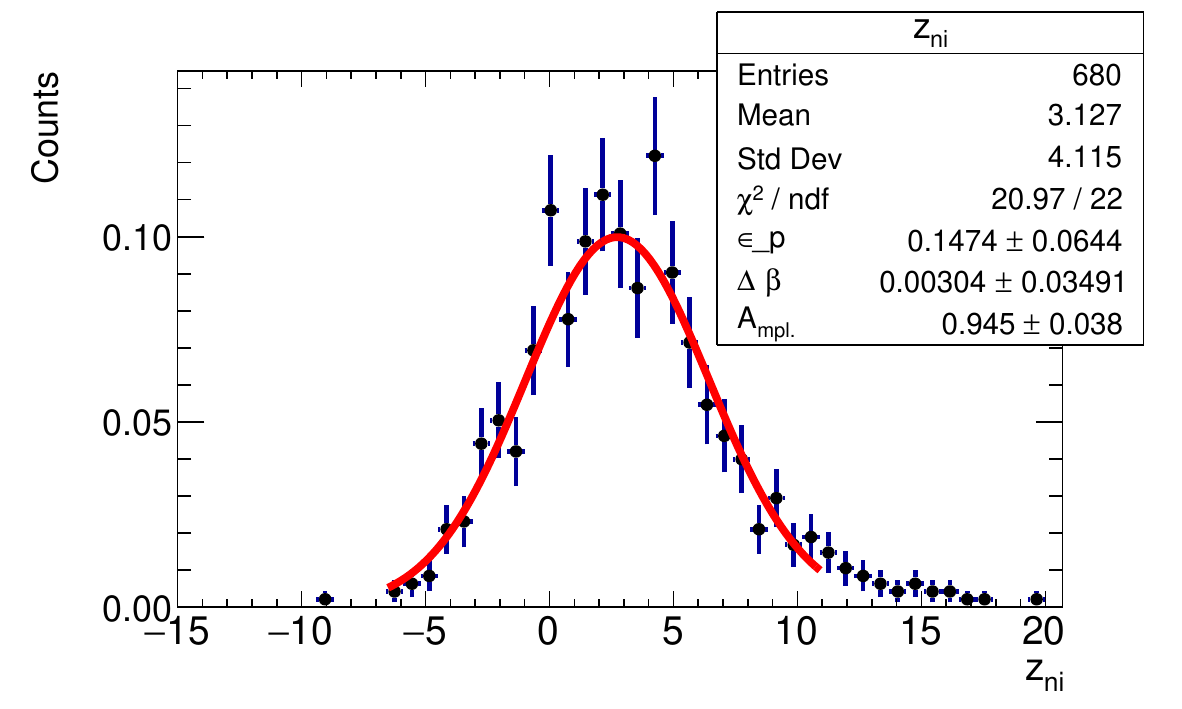}
\includegraphics[width=0.49\textwidth, height=\h\textwidth]{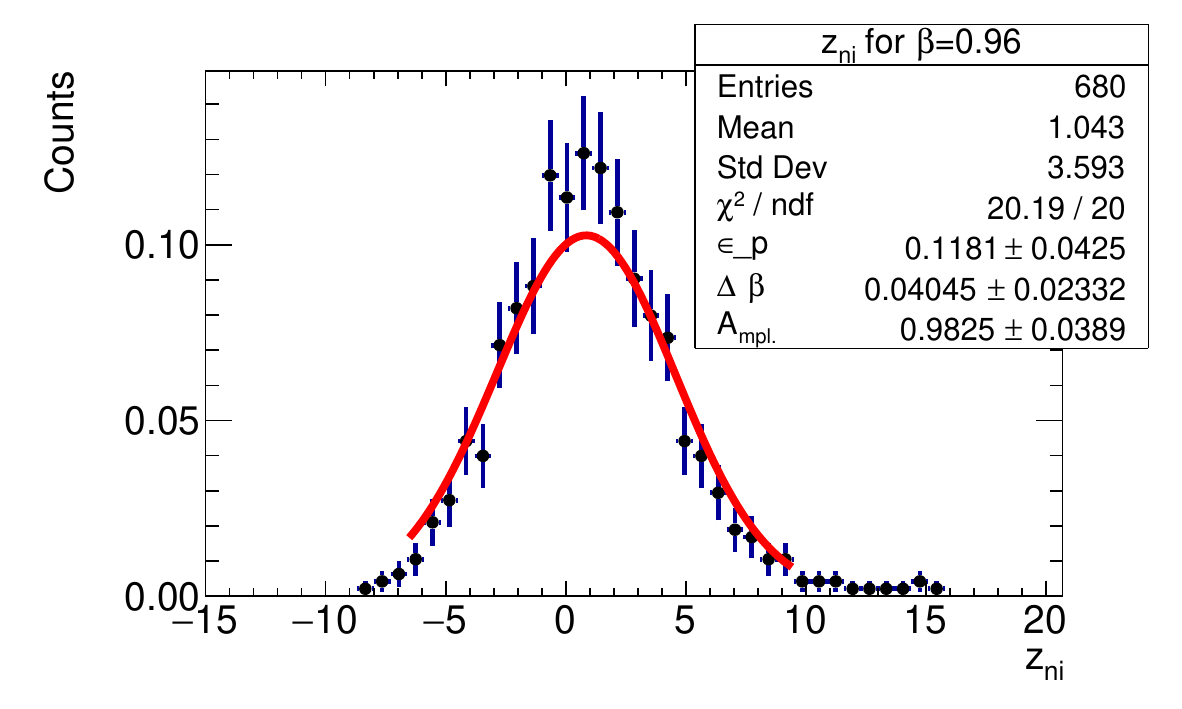}
\caption{(Left): The results of fitting the function given by \cref{fit-final} to the normalized $z_{ni}$ histogram shown in \cref{fig:z-distribution} (mockdataset) with $\sigma(z_k)/\text{VEM}=\{ 4.11,3.40,4.02,3.91\}$ and $f_k=\{0.15,0.38,0.46,01\}$ and $\avg{\bar{S}^\upmu}_k/\text{VEM}=\{15.57,17.25,19.37,21.61\}$ and with the two-parameters exponential scaling model.
The four individual $z_k$-distributions  match the $z_{ni}$-histogram for $\eps_\text{p}=0.147\pm0.062$, $\Delta\beta=0.003\pm0.035$ and scaling parameter $A_\text{mpl}=0.945\pm0.038$. (Right): The four individual $z_k$-distributions  match $z_{ni}$-histogram for mockdataset with $\beta=0.96$ and for $\eps_\text{p}=0.118\pm0.043$, $\Delta\beta=0.040\pm0.023$ and $A_\text{mpl}=0.983\pm0.039$.}
\label{fig:mockdata-solutions}
\end{figure}
\vspace{-0.3cm}
\section{Results of the fit of four individual $z_k$-distributions to $z_{ni}$-histogram }
\vspace{-0.2cm}
The results are shown in \cref{fig:mockdata-solutions} and \cref{tab-mock}.
We see that the fit can reproduce the ratio of the muon signals of simulations using \textsc{Epos-LHC} (mock data) and \textsc{QGSJetII-04} within ${\sim}5\%$: as we already previously mentioned, the ratio for MC-true is $r^\text{MC}_\text{true}=1.10\pm0.04$ and the average reconstructed ratio (from \cref{tab-mock}) is $1.15\pm0.06$.
The difference is caused by the fact that the signal for the mock dataset is not exactly equal to the one for \textsc{Epos-LHC} (\cref{tab-mock}).
We also recover the $\beta$ parameter (average ${\simeq}0.92$) for the studied set, because parameter $\Delta\beta$ is zero within its error i.e.\ $\Delta\beta=0.003\pm0.035$.
Finally we can check our solution by comparing the mean given by \cref{z_bar} and this from a Gaussian fit to the $z$-histogram shown in \cref{fig:z-distribution}.
We get $2.74\pm0.49$\,VEM vs.\ $\avg{z_{ni}}=2.83\pm0.16$\,VEM, which agree very well within the limits. 
We have a standard deviation match by definition, because $\sigma^2(z_{ni})=\int \sum_kf_k z^2_{nik} \,P_k(z_k,\sigma(z_k))\,\dd z_{nik}=\sum_kf_k\sigma^2(z_k)$.

\begin{table*}
\small
\centering
\begin{tabular}{lcccccccc}
\toprule
$k$ & $r_{\upmu,k}$ & $\avg{\bar{S}^\upmu_k}/\text{VEM}$ & $\avg{S_{\upmu,k}^\text{rec}}/\text{VEM}$  & $\delta$ & $r_{\upmu,k}$& $\avg{S_{\upmu,k}^\text{rec}}/\text{VEM}$& $\delta$\\
\midrule
p & $1.16 \pm 0.06$ & $15.57 \pm 0.17$ & $18.03 \pm 0.18$ & 4.2\% & $1.13\pm0.04$ &$17.52\pm0.17$ &1.0\%\\     
He & $1.15 \pm 0.06$ & $17.25 \pm 0.19$ & $19.90 \pm 0.20$ & 4.3\% &$1.07\pm0.07$ & $18.11\pm0.30$&1.4\%\\
N & $1.15 \pm 0.06$ & $19.37 \pm 0.20$ & $22.26 \pm 0.21$ &  5.3\% &$1.01\pm0.09$ &$19.24\pm0.38$ &1.9\%\\
Fe & $1.14 \pm 0.06$ & $21.61 \pm 0.23$ & $24.73 \pm 0.24$ & 5.6\%&$0.96\pm0.10$ & $20.23\pm 0.25$ &2.4\%\\
\bottomrule
\end{tabular}
\caption{Values of the muon rescaling factors obtained with the fitting procedure, and of the MC muon signal, the reconstructed muon signals, for all primaries considered and with $f_\text{p}=0.15$, $f_\text{He}=0.38$, $f_\text{N}=0.46$, and $f_\text{Fe}=0.01$.
The overestimation $\delta=(\avg{S_{\upmu,i}^\text{rec}} - \avg{S_{\upmu,i}^\text{mock}}) / \avg{S_{\upmu,i}^\text{mock}}$ of the reconstructed muon signal compared to the one from the mock dataset is also provided.
The errors shown in the  four column are the square root of the sum of the squares of the errors $\delta r_{\upmu,k}$ and $\delta\avg{\bar{S}_{\upmu,k}^\text{rec}}$, i.e.\ those listed in the second and third columns, respectively. The last free columns show results for mockdataset with $\beta=0.96$.}
\label{tab-mock}
\end{table*}

Since the true value of the hybrid dataset may differ from that of the hadron interaction models used in this analysis, it would be interesting to perform the same analysis for a sample dataset built from the \textsc{Epos-LHC} sample, but with hadron interaction evolution. For a sample dataset built from the \textsc{Epos-LHC} sample, we constructed a mockdataset  with the evolution of the  mean muon signal as a function of primary mass, leading to a significantly different exponent value $\beta\simeq0.96$.
This allows us to investigate whether the fitting procedure is able to recover this value as well.
The average muon signal of the new sample set as a function of primary mass is shown in~\cref{tab-mock}.
Two features of this  mock dataset can be noticed: that for nitrogen, a slight rescaling from \textsc{QGSJetII-04} to the mock dataset is needed ($r_{\upmu,\text{N}}=1.01$) and that for iron the average muon signal is lower than 1 for the mock dataset ($r_{\upmu,\text{Fe}}=0.96$).
The results of the fit are shown in \cref{fig:mockdata-solutions} (right) and in \cref{tab-mock}.
We can see that the negative scaling of the primary iron is slightly underestimated, while the signal is well recovered for all other elements.
The muon signal from the mocked-up dataset is recovered within 2.4\%.
Moreover, the fitting of the reconstructed muon signal gives a value of $\Delta\beta=0.04$ which
agrees quite well with the expectation $\beta=0.955\pm0.005$, although error of the $\Delta\beta$ is quite large 0.02.

\vspace{-0.3cm}
\section{Summary and Conclusion}
\label{conclusion}
\vspace{-0.2cm}
The method presented in this work recovers the mean muon signal and provides the ability to calculate muon signals for each element in the considered sample of real-like events.
In this work, we have been performed calculations of muon scaling factors and $\beta$ exponents, by  fitting  a four-element Gaussian distribution to the overall z-histogram, with two-parameter scaling model which should follow Heitler-Matthews progression.  This work shows that the $z$-method can be applied to hybrid events to determine the muon signal, the scaling factor (total and for each element), and the $\beta$ exponent.

{\bf Acknowledgments:} The authors are very grateful to the Pierre Auger Collaboration for providing the tools necessary for simulation
for this contribution. The authors would like to thank the colleagues from the Pierre Auger Collaboration for
all the fruitful discussions~\footnote{We want to acknowledge the support in Poland from the National Science Centre grant
No.~2016\-/\-23\-/\-B\-/\-ST9\-/\-01635, grant No.~2020/39/B/ST9/01398, grant No.~2022/45/B/ST9/02163 and from the Ministry of Education and
Science grant No.~DIR/WK/2018/11 and grant No.~2022/WK/12.}.

%% Full authors list (ONLY FOR COLLABORATIONS)
%\clearpage
%\section*{Full Authors List: \Coll\ Collaboration}
%
%\noindent \textbf{Note comment afterwards:} Collaborations have the possibility to provide an authors list in xml format which will be used while generating the DOI entries making the full authors list searchable in databases like Inspire HEP. For instructions please go to icrc2021.desy.de/proceedings or contact us under icrc2021proc@desy.de.\\
%
%\scriptsize
%\noindent
%first.author$^1$, 
%second.author$^2$, 
%third.author$^3$ % .... more names
%and 
%last.author$^{n}$ \\
%
%\noindent
%$^1$first.affiliation.
%$^2$second.affiliation. % .... more affiliation
%$^{m}$last.affiliation.

\end{document}